# Enhanced Medium Range Order in Vapor Deposited Germania Glasses at Elevated Temperatures

Le Yang,[1]* Gabriele Vajente,[2] Mariana Fazio[3], Alena Ananyeva[2], GariLynn Billingsley[2], Ashot Markosyan[4], Riccardo Bassiri[4], Kiran Prasai[4], Martin M. Fejer[4], Martin Chicoine[5], François Schiettekatte[5], Carmen S. Menoni[1,3]*

[1]Department of Chemistry, Colorado State University, Fort Collins, CO 80523, USA.
[2]LIGO Laboratory, California Institute of Technology, Pasadena, CA 91125, USA.
[3]Department of Electrical and Computer Engineering, Colorado State University, Fort Collins, CO 80523, USA.
[4]Edward L. Ginzton Laboratory, Stanford University, Stanford, CA 94305, USA
[5]Départment de physique, Université de Montréal, Québec H3C 3J7, Canada.

yangle@colostate.edu, carmen.menoni@colostate.edu

## Abstract

Glasses are nonequilibrium solids with properties highly dependent on their method of preparation. In vapor-deposited molecular glasses, structural organization could be readily tuned with deposition rate and substrate temperature. Herein, we show the atomic arrangement of strong network forming $GeO_2$ glass is modified at medium range (< 2 nm) through vapor deposition at elevated temperatures. Raman spectral signatures distinctively show the population of 6-membered $GeO_4$ rings increases at elevated substrate temperatures. Deposition near the glass transition temperature is more efficient than post-growth annealing in modifying atomic structure at medium range. The enhanced medium range organization correlates with reduction of the room temperature internal friction. Identifying the microscopic origin of room temperature internal friction in amorphous oxides is paramount to design the next generation interference coatings for mirrors of the end test masses of gravitational wave interferometers, in which the room temperature internal friction is a main source of noise limiting their sensitivity.

## Teaser

The medium range order in $GeO_2$ thin film glasses is higher with elevated deposition temperature.

## Introduction

Glasses are nonequilibrium, non-crystalline materials that retain in their structure at short and medium range (< 2 nm) information about the history of preparation (1, 2). For melt-quenched glasses, a slow cooling rate towards the glass transition temperature ($T_g$) allows for an adequate configurational sampling that drives the system to lower energy states in the potential energy landscape (PEL). Physical vapor deposition (PVD) is an efficient means of rapid cooling to produce glassy materials. Altering the deposition conditions, such as substrate temperature ($T_{sub}$) and deposition rate, makes it possible to manipulate the atomic ordering that in turn shapes the properties of the vapor-deposited glasses (3-6). For example, in vapor-deposited thin films of itraconazole, a glass forming smectic liquid crystal, the orientation of the molecule's long axis tends to align with the surface normal when the deposition rate is around 0.2 Å/s, whereas nearly isotropic orientation is preferred when the deposition rate is 3 orders of magnitude higher (7). In a similar fashion, the substrate temperature plays a role in affecting the molecular packing in thin-film organic glasses. Depositing *N,N'*-bis(3-methylphenyl)- *N,N'*-diphenylbenzidine at around 0.8 $T_g$, produces glasses with a strong horizontal orientation, while at 0.95 $T_g$ weak vertical orientation is observed (3, 8). *In silico* vapor deposition of glasses predicts that when depositing at the Kauzmann temperature, where the configurational entropy vanishes, it would be possible to achieve a uniform structural configuration characteristic of ultrastable glassy materials (9). It remains to be answered experimentally whether or not the atomic arrangement of strong network-forming glasses, such as amorphous $SiO_2$ (a-$SiO_2$) and $GeO_2$ (a-$GeO_2$), could be modified by altering the substrate temperature since the restructuring of strong covalent bonds is involved.

A question that follows is whether elevated temperature deposition of a-$GeO_2$ would result in atomic rearrangements that alter the distribution of two-level systems (TLSs) in glassy materials. TLSs in the PEL model are used to describe the acoustic and thermal properties of amorphous solids at low temperatures (10). They are represented by asymmetric double-well potentials in some configuration coordinate. Transitions between the wells, which are thermally activated at temperatures above ~5 K, and quantum tunneling dominated at lower temperatures, are associated with the rearrangement of a small group of atoms at temperatures well below $T_g$ (11). Recent insights into the properties of vapor-deposited glasses show that TLSs could possibly be drastically reduced at selected deposition conditions (2). In indomethacin thin film glasses grown at 0.85 $T_g$, remarkable suppression of TLSs was found due to the particular molecular arrangement influenced by the deposition conditions (12). For network forming glasses such as a-Si, the density of TLSs has been shown to be reduced by at least one order of magnitude when $T_{sub}$ increased from 473 K to 673 K (13, 14). This behavior has been ascribed to a more ordered amorphous network achieved with elevated temperature deposition. Motivated by these results, the gravitational wave community has explored the possibility to lower the density of TLSs through elevated temperature deposition of a-$Ta_2O_5$ as a way to reduce room temperature internal friction in the coatings of the end test mass mirrors (15, 16). Internal friction in amorphous materials is generally framed in terms of energy coupling from an elastic field into TLSs. The coupling causes excitations visualized as transitions between the two wells in a TLS. The relaxation of these excitations through various mechanisms characterizes the system's internal friction (17, 18). Room temperature internal friction, in accord with the fluctuation-dissipation theorem (19), leads to thermally driven fluctuations in the amorphous coatings that limit the sensitivity of gravitational wave detectors (20-22). Recent work in a-Si thin films showed room temperature internal friction reduced from around $2 \times 10^{-4}$ to $0.5 \times 10^{-4}$ when $T_{sub}$ increased from 293 K to 673 K (~ 0.80 $T_g$) (23, 24). Nevertheless, a correlation between structural organization and room temperature internal friction for strong network forming glasses deposited at elevated substrate temperature is still lacking.

Herein, we describe the modifications in the atomic configuration at medium range of a-$GeO_2$ thin films vapor deposited at elevated substrate temperatures. The lower $T_g$ of $GeO_2$, around 788 K (25), in comparison with $T_g$ around 1475 K for $SiO_2$ (26), makes the deposition at $T_{sub}$ near $T_g$ accessible. The signatures of structural ordering, obtained from Raman spectroscopy, distinctively show the arrangement of $GeO_4$ tetrahedra into 6-membered rings increases when depositing at elevated substrate temperatures. It is also demonstrated that the deposition near $T_g$ is more efficient than post-growth annealing in modifying the atomic structure at medium range. These structural modifications correlate with the room temperature internal friction of a-$GeO_2$ thin films, which decreases by as much as 44% when the film is deposited at 0.83 $T_g$. In combination, the results demonstrate a strong correlation between medium range order and room temperature internal friction as predicted by theory.

**Results**

The physical vapor deposition of amorphous oxides is characterized by hit and stick processes in which the atomic relaxation and formation of a more stable structure are constrained when the substrate temperature is low. Elevated $T_{sub}$ and post-deposition annealing alter the organizational state of the deposited glasses by enabling the system to explore nearby lower minima in the energy landscape through atomic rearrangements. For a-$GeO_2$ that has strong directional covalent bonding, the most dominant structural order is at the medium range, which can be defined in terms of the connection of the $GeO_4$ tetrahedra (27-29). The (Ge-O-Ge) connection chain has a ring shape with a size that is determined by the number of Ge atoms within the closed path. Rings of various sizes ranging from 3 to 10 with a maximum distribution of around 6-membered rings are predicted by models of a-$GeO_2$ (30). This structural information can be obtained from x-ray and neutron diffraction in combination with modeling (30, 31) or alternatively from Raman spectroscopy that is sensitive to local vibrations at the medium range in glasses.

The Raman spectra of a-$GeO_2$ thin films, shown in Fig. 1, are characterized by strong peaks at around 430 $cm^{-1}$ and 510 $cm^{-1}$. The former corresponds to the symmetric stretching of bridging oxygen in 6-membered rings ($A_6$) (32). The latter corresponds to the oxygen-breathing mode associated with 3-membered rings ($A_3$) (33, 34). Two Raman bands at 560 $cm^{-1}$ and 595 $cm^{-1}$ assigned to the transverse optical and longitudinal optical asymmetric stretching of bridging oxygen, respectively (Fig. S1), overlap with $A_3$. This overlap introduces significant uncertainty in the fitting of the $A_3$ peak at 510 $cm^{-1}$. The integrated $A_6$ area was normalized to the total integrated area of the Raman spectrum between 200 to 700 $cm^{-1}$ to evaluate the change in the population of 6-membered rings. Figure 1 qualitatively contrasts the difference in the ring distribution of a-$GeO_2$ samples deposited at $T_{sub}$ = room temperature and $T_{sub}$ = 0.83 $T_g$, which indicates that high temperature deposition favors a larger fraction of 6-membered rings, i.e., a more ordered structure, as described below.

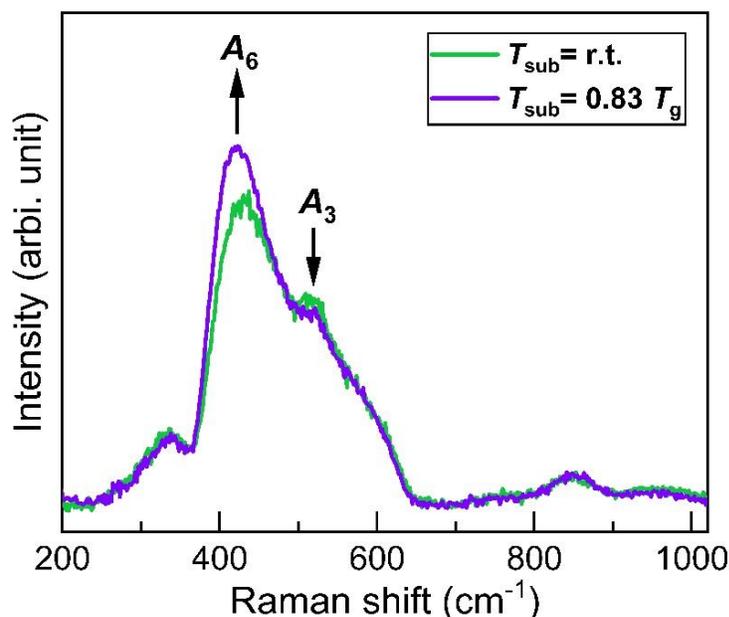

**Fig.1. Raman scattering spectra of a-GeO$_2$ thin films deposited at different temperatures.** $T_{sub}$ = room temperature (green) and $T_{sub}$ = 0.83 $T_g$ (purple). The $A_6$ peak at around 430 cm$^{-1}$ corresponds to the symmetric stretching of bridging oxygen in 6-membered rings. The $A_3$ peak at around 510 cm$^{-1}$ corresponds to the breathing motion of bridging oxygen in 3-membered rings. The peak at 337 cm$^{-1}$ is assigned to the Ge 'deformation' motion within the network (33, 34).

We focus first on the evolution of the ring distribution in a-GeO$_2$ deposited at $T_{sub}$ = room temperature as the system approaches lower energy states with post-deposition annealing, visualized as changes in normalized $A_6$ in Fig. 2. $A_6$ does not significantly vary after the first annealing step at an annealing temperature $T_{an}$ = 573 K. $A_6$ shows a pronounced increase of 29% to 0.63 ± 0.04 after annealing at $T_{an}$ = 623 K. Annealing to higher temperature continues to relax the atomic structure, leading to an increase of $A_6$ to 0.72 ± 0.03. Overall, the trend indicates that the population of large 6-membered rings increases. The breakup of small (three- or four-membered) rings, qualitatively identified by a reduction in the $A_3$ peak intensity, could be attributed to heavily strained inter-tetrahedral bridging bonds (35, 36). The increase in the population of 6-membered rings indicates that an increased medium range order is achieved in a-GeO$_2$ thin films upon annealing.

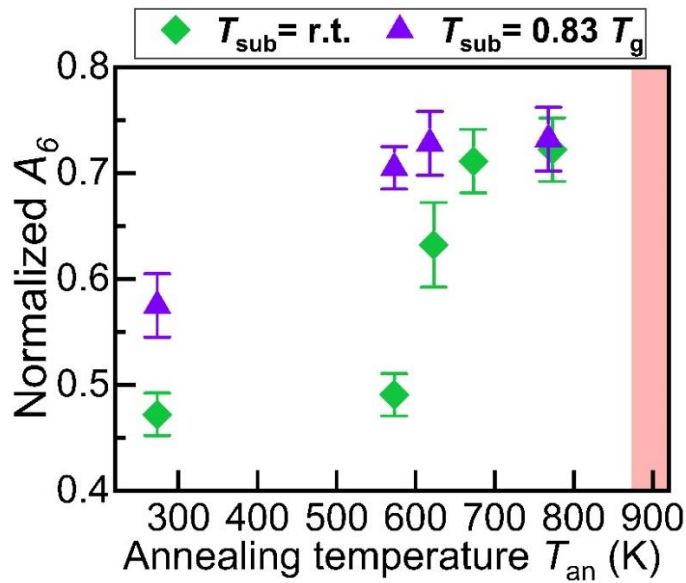

**Fig.2. Structural relaxation in a-GeO$_2$ thin films identified by changes in normalized $A_6$ versus annealing temperature $T_{an}$.** Samples deposited at $T_{sub}$ = room temperature and $T_{sub}$ = 0.83 $T_g$ are represented by green rhomboids and purple triangles, respectively. The temperature on each annealing step was held for 10 hours. The red shaded region in the figure indicates the start of crystallization at around 873 K for all samples.

Deposition at elevated $T_{sub}$ brings more significant changes to the ring distribution in a-GeO$_2$, as shown in Fig. 2. $A_6$ is 0.58 ± 0.03 for the sample deposited at $T_{sub}$ = 0.83 $T_g$, and 0.47 ± 0.02 for the sample deposited at $T_{sub}$ = room temperature. The larger fraction of 6-membered rings confirms the formation of a more ordered atomic structure at elevated $T_{sub}$. Moreover, the results show that the annealing temperature required to achieve a high medium range order before crystallization, defined by $A_6$ ~ 0.71, is lower at $T_{an}$ = 573 K, when a-GeO$_2$ is deposited at $T_{sub}$ = 0.83 $T_g$, in comparison to $T_{an}$ = 673 K for the a-GeO$_2$ deposited at $T_{sub}$ = room temperature. Further annealing of the sample deposited at $T_{sub}$ = 0.83 $T_g$ to $T_{an}$ = 773 K resulted in increased normalized $A_6$ ~ 0.72. All samples remained amorphous after annealing at $T_{an}$ = 773 K for 10 hours (Figs. S2-S3). During room temperature deposition, incident sputtered particles rapidly lose their energy to the substrate before exploring the entire configuration space (37), whereas elevated substrate temperature promotes a larger sampling that leads to a higher degree of organization. It is through elevated substrate temperature that vapor-deposited glasses are able to achieve a more ordered structural organization (9, 38, 39).

The structural rearrangements at medium range play a role in lowering the potential energy of the glass system and are strongly dependent on the substrate temperature, as recently shown in molecular dynamic simulations of vapor-deposited a-SiO$_2$ films (39). In comparison to a-SiO$_2$ films deposited at $T_{sub}$ well below $T_g$, the films with lower potential energy prepared at the optimal $T_{sub}$ near $T_g$ have a higher fraction and a narrower distribution of rings centered at 6-membered rings, suggesting a greater structural uniformity.

The changes in the medium range order of a-GeO$_2$ and their impact in modifying thermally activated TLSs are assessed from the room temperature internal friction Q$^{-1}$ of thin films deposited at different $T_{sub}$ and post-deposition annealed. Q$^{-1}$ will refer to internal friction at room temperature unless otherwise noted. Previous modeling and experimental results of amorphous oxides, such as a-$Ta_2O_5$, have suggested that an increased medium range order correlates with the reduction in $Q^{-1}$ (40, 41) at room temperature. The evolution of $Q^{-1}$ for a-$GeO_2$ deposited at $T_{sub}$ = room temperature with annealing is shown in Fig. 3A. The as-prepared a-$GeO_2$ thin film has $Q^{-1}$ of $(2.98 \pm 0.27) \times 10^{-4}$, which reduces to $(2.56 \pm 0.52) \times 10^{-4}$ after the first annealing step at $T_{an}$ = 573 K. A significant decrease by 49% is obtained after the $T_{an}$ = 623 K annealing step. Notably, the annealing temperature after which $Q^{-1}$ undergoes a sharp decrease, coincides with the turning point in the increase of normalized $A_6$ in Fig. 2. Beyond $T_{an}$ = 673 K, $Q^{-1}$ plateaus at a value of $(1.00 \pm 0.13) \times 10^{-4}$.

Comparison of $Q^{-1}$ for as-prepared a-$GeO_2$ deposited at $T_{sub}$ = room temperature, $T_{sub}$ = 0.60 $T_g$, and $T_{sub}$ = 0.83 $T_g$ shows a steady decrease with an increase in $T_{sub}$. The a-$GeO_2$ sample deposited at $T_{sub}$ =0.83 $T_g$ has the lowest $Q^{-1}$ = $(1.66 \pm 0.14) \times 10^{-4}$, which is 44% less than $Q^{-1}$ of the sample deposited at room temperature. Figure 3 shows a reduction in $Q^{-1}$ with $T_{an}$ for all samples, although the rate at which $Q^{-1}$ reaches its minimum value is different for each one. The a-$GeO_2$ thin film prepared at $T_{sub}$ =0.60 $T_g$ reaches $Q^{-1}$ =1.00 $\times 10^{-4}$ after annealing at $T_{an}$ = 623 K, yet the one prepared at $T_{sub}$ = 0.83 $T_g$ achieves this $Q^{-1}$ after annealing at $T_{an}$ = 573 K. Annealing beyond $T_{an}$ = 623 K does not decrease $Q^{-1}$ below 1.00 $\times 10^{-4}$ for the high temperature deposited samples. It is also worth noting that the a-$GeO_2$ thin film deposited at $T_{sub}$ = 0.60 $T_g$ (473 K) without thermal treatment has $Q^{-1}$ = $(2.39 \pm 0.15) \times 10^{-4}$, which is comparable to $Q^{-1}$ = $(2.56 \pm 0.52) \times 10^{-4}$ for the sample deposited at room temperature and annealed at $T_{an}$ = 573 K for 10 hours. Considering that the same level $Q^{-1}$ is obtained at a lower $T_{an}$ during a significantly shorter deposition time of around 1 hour compared to annealing for 10 hours, it indicates a higher efficiency of elevated temperature deposition over annealing for reducing $Q^{-1}$ of a-$GeO_2$ thin films.

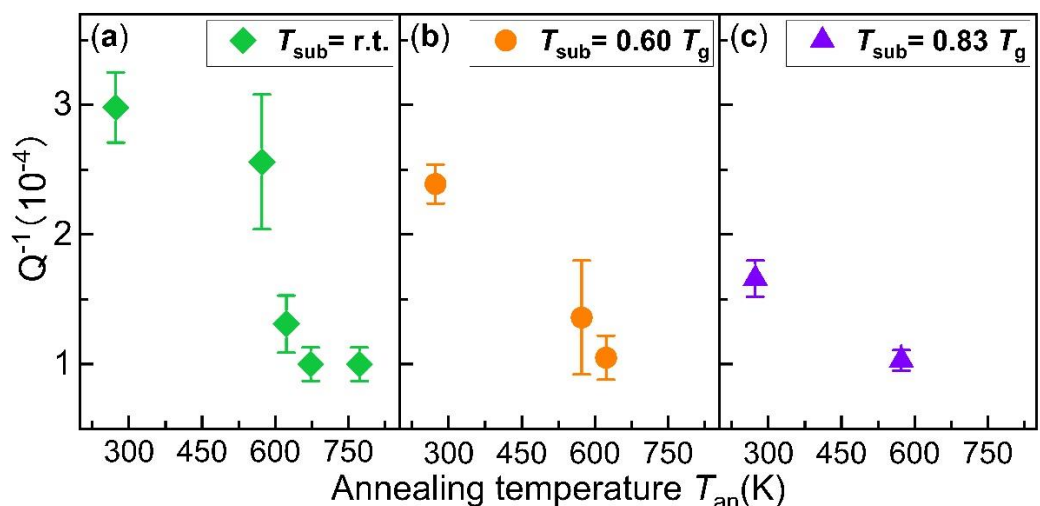

**Fig.3. Room temperature internal friction $Q^{-1}$ of a-$GeO_2$ thin films deposited at different temperatures.** (**A**) $T_{sub}$ = room temperature (green), (**B**) $T_{sub}$ = 0.60 $T_g$ (orange), and (**C**) $T_{sub}$ = 0.83 $T_g$ (purple). The annealing temperature to reach the lowest room temperature internal friction is $T_{an}$ = 673 K, $T_{an}$ = 623 K, and $T_{an}$ = 573 K for samples deposited at $T_{sub}$ = room temperature (green), $T_{sub}$ = 0.60 $T_g$ (orange), and $T_{sub}$ = 0.83 $T_g$ (purple), respectively.

The inverse relationship of $Q^{-1}$ with normalized $A_6$ in Fig. 4 demonstrates a strong and direct correlation between lowering $Q^{-1}$ and increasing medium range order characterized by a larger fraction of 6-membered rings in a-$GeO_2$. Such observation echoes what has been found for vapor-deposited a-$SiO_2$ (42), where a steady decrease in the population of 3-membered rings is linked to the reduction in $Q^{-1}$ during extended annealing. Similar behavior has also been observed in Ref (43), in which analyses of grazing-incidence pair distribution functions of $ZrO_2$-doped $Ta_2O_5$ reveal a systematic change in the medium range order with annealing temperature. Modeling shows that the atomic rearrangements that occur at medium range involve a decrease in the population of edge- and face-sharing polyhedra and an increase in corner-sharing ones, which agree with an expansion in the polyhedral ring connection. These modifications correlate with a steady decrease in $Q^{-1}$ at room temperature. Further evidence directly linking the structural changes with TLSs for $ZrO_2$-doped $Ta_2O_5$ is obtained from Ref (44). In this work, the energy landscape of $ZrO_2$-doped $Ta_2O_5$ was explored by searching for TLSs using molecular dynamic simulations. It is found that the TLSs can be sorted into two types depending on whether the cation-oxygen bond within the polyhedron breaks, namely cage-breaking and non-cage breaking transitions. The simulations show that a significant number of TLSs from cage-breaking events are responsible for room temperature $Q^{-1}$ in the amorphous oxide. Elimination of such transitions by expanding the polyhedral connections, i.e., increasing the ring size, would thus result in lower $Q^{-1}$. In combination, these theories provide a consistent interpretation of the correlation between $Q^{-1}$ and medium range order observed in a-$GeO_2$ thin films.

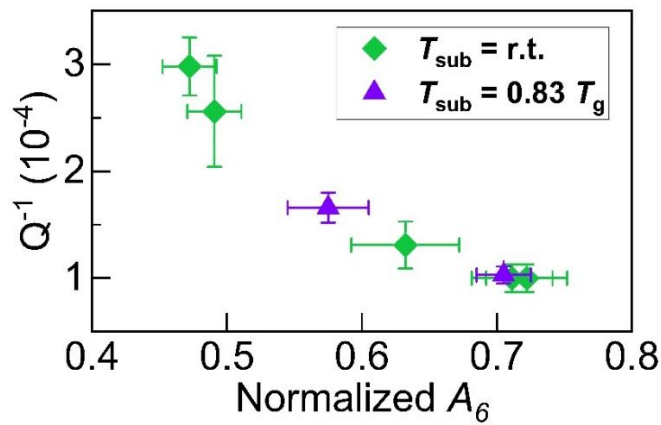

**Fig.4. Room temperature internal friction $Q^{-1}$ as a function of normalized $A_6$ for a-$GeO_2$ thin films deposited at different $T_{sub}$ and subsequently annealed.** Samples deposited at $T_{sub}$ = room temperature and $T_{sub} = 0.83\ T_g$ are represented by green and purple marks, respectively.

**Discussion**

The results conclusively show the atomic organization of strong network forming a-$GeO_2$ can be modified with elevated temperature deposition, similar to the behavior observed in organic glasses (3-8). The changes in the atomic arrangement are characterized by an increase in the fraction of 6-membered rings. This structural reorganization corresponds to an increase in medium range order, which is more significant when a-$GeO_2$ is deposited at $T_{sub} = 0.83\ T_g$ compared to films deposited at $T_{sub}$ = room temperature and subsequently annealed.

Furthermore, the results establish a strong correlation between medium range order and room temperature internal friction in a-$GeO_2$. The more ordered atomic arrangement leads to a reduction in room temperature internal friction of as much as 44% when the film is deposited at $T_{sub} = 0.83\ T_g$ compared to $T_{sub}$ = room temperature. Under optimum deposition and annealing conditions, the room temperature internal friction of a-GeO$_2$ reaches a value of $Q^{-1} = 1.00 \times 10^{-4}$, which is among the lowest in amorphous oxides and comparable to vapor deposited a-SiO$_2$ (42). These findings are relevant for identifying suitable amorphous oxide candidates for mirror coatings of the end-test masses of gravitational wave interferometers, in which the coating's room temperature internal friction is the main source of noise limiting their sensitivity.

**Materials and Methods**

a-GeO$_2$ thin films were deposited with a 4Wave LANS physical vapor deposition system (45, 46). A gridless ion source was used to generate low energy Ar ions that were accelerated towards a high purity Ge target negatively biased at 800 V. This created a sputter plume that deposited onto the substrate. Oxidation was achieved by flowing 6 sccm of oxygen near the substrate surface. For high temperature deposition, a heating lamp was positioned under the rotating substrate stage. The stage temperature was increased to the set point 10 minutes before deposition for equilibration purposes and maintained through the process. The rate for room temperature and high temperature deposition was kept at 1.1 and 1.2 Å/s, respectively. The use of approximately the same deposition rate ruled out the influence of this parameter in atomic reorganization. Thermal annealing of each sample after deposition was carried out at ambient conditions by ramping up the temperature by 100 K/hour to $T_{an}$, holding at $T_{an}$ for 10 hours, and ramping down at 100 K/hour to room temperature.

Raman scattering was performed with a Horiba LabRAM HR Evolution Spectrometer. A frequency doubled Nd:YAG laser of 532 nm wavelength and 10 mW average power was used for excitation. The laser beam was focused onto the sample's surface using a ×100 objective. Three spectra of 60 seconds acquisition time were collected for each sample and averaged to improve the signal to noise. Deconvolution of the peaks in the Raman spectrum was carried out by fitting the peaks with Gaussian lineshapes. The Raman spectrum contains a peak at 337 cm$^{-1}$ which corresponds to the Ge motion within the network; $A_6$ that corresponds to 6-membered rings at 430 cm$^{-1}$ (Table S1); and the broad feature composed of the dominant $A_3$ peak at 510 cm$^{-1}$ associated with 3-membered rings and peaks at 560 cm$^{-1}$ and 595 cm$^{-1}$ associated with the Ge-O-Ge bending modes. In the fitting, the FWHM of $A_3$ was constrained to 75 cm$^{-1}$ (47). The overlap of $A_3$ with the peaks at 560 cm$^{-1}$ and 595 cm$^{-1}$ produces a high uncertainty in the calculation of the $A_3$ area. It is for this reason that we use the area of $A_6$ normalized to the area of the spectrum from 200 cm$^{-1}$ to 700 cm$^{-1}$ to represent changes in the population of 6-membered rings.

Room temperature internal friction ($Q^{-1}$) of the a-GeO$_2$ thin films was measured with a ring down system (48, 49). The ~ 500-nm-thick a-GeO$_2$ samples for this purpose were deposited on high quality fused silica disk of 75-mm diameter and 1-mm thickness. A gentle nodal suspension method was used to support the sample inside a vacuum chamber held at a pressure below 10$^{-6}$ Torr. After exciting the resonant mode, the oscillation amplitude at each frequency $f_i$ was tracked to obtain the decay time $\tau_i$. The room temperature internal friction $Q_i^{-1}$ for each mode was then computed through the following relation,

$$Q_i^{-1} = 1/(\pi f_i \tau_i)$$

At each frequency, 8 measurements were performed to obtain an averaged $Q_i^{-1}$. A mean value of $Q^{-1}$ for each sample was then obtained by averaging the $Q_i^{-1}$ over a frequency range of 1-20 kHz.

Rutherford backscattering spectrometry (RBS) analyses were performed using He$^+$ ions at 2.035 MeV with a 1.7 MV Tandetron accelerator. Spectra were acquired with samples on Si substrates tilted at 7° to minimize channeling. The scattered ions were collected at an angle of 10° from the beam. The experimental spectra were simulated using the SIMNRA software (50) in order to extract the composition and areal atomic density of the deposited layers. All samples were stoichiometric after annealing at 773 K (Table S2).

**Acknowledgments**

The Raman scattering was performed at the Raman Microspectroscopy Laboratory in the Department of Geological Science at the University of Colorado-Boulder.

**Funding:**
National Science Foundation LIGO program through grants No. 1710957 (LY, CSM) and 1708010 (MF).
NSF awards PHY-1707866, PHY-1708175 (AM, RB, MMF)
GBMF Grant No. 6793 (AM, RB, MMF)
FRQNT through the Regroupement québécois sur les matériaux de pointe (RQMP)
Natural Sciences and Engineering Research Council of Canada (MC, FS)

**Author contributions:**
LY, MF, CSM conceived the experiment.
LY, MF performed the thin film deposition and material spectroscopy analysis.
GV, AA, GB performed the room temperature internal friction measurements.
AM, RB, KP, MMF contributed to the XRD measurements.
MC, FS contributed to the RBS measurements.
All authors contributed to the preparation of the manuscript.

**Competing interests:** Authors declare that they have no competing interests.

**Data and materials availability:** All data are available in the main text and/or the Supplementary Materials.